\documentclass[11pt,twoside]{article}
\usepackage{asp2004}
\usepackage{psfig}
\usepackage{epsf}
\usepackage{graphics}
\usepackage{lscape}
\begin{document}

\title{Black Holes in Massive Star Clusters}

\author{Steve McMillan}
\affil{Department of Physics, Drexel University, Philadelphia,
	PA 19104, U.S.A.}

\author{Holger Baumgardt}
\affil{Institute of Advanced Physical and Chemical Research (RIKEN),
	2-1 Hirosawa, Wako-shi, Saitama 351-019, Japan}

\author{Simon Portegies Zwart}
\affil{Astronomical Institute `Anton Pannekoek' and Institute for
	Computer Science, University of Amsterdam, Kruislaan 403,
	the Netherlands}

\author{Piet Hut}
\affil{School of Interdisciplinary Studies, Institute for Advanced
	Study, Princeton, NJ 08540, U.S.A.}

\author{Junichiro Makino}
\affil{Department of Astronomy, University of Tokyo, Tokyo 113, Japan}

%---------------------------------------------------------------------

\begin{abstract}
Close encounters and physical collisions between stars in young dense
clusters can result in new channels for stellar evolution, and may
lead to the formation of very massive stars and black holes via
runaway merging.  We present some details of this process, using the
results of N-body simulations and simple analytical estimates to place
limits on the cluster parameters for which it expected to occur.  For
small clusters, the mass of the runaway is effectively limited by the
total number of high-mass stars in the system.  For larger clusters,
the runaway mass is determined by the fraction of stars that can
mass-segregate to the cluster core while still on the main sequence.
In typical cases, the result is in the range commonly cited for
intermediate-mass black holes.  This mechanism may therefore have
important implications for the formation of massive black holes and
black-hole binaries in dense cluster cores.
\end{abstract}

\thispagestyle{plain}

%---------------------------------------------------------------------

\section{Introduction: Young Dense Clusters in the Galaxy and Beyond}

Among the many massive young clusters now known throughout the local
universe, perhaps the most interesting to dynamicists are those in
which stellar dynamical time scales are short enough that the cluster
can undergo significant structural change during the lifetimes of the
most massive stars.  In such clusters, dynamical evolution opens up
novel avenues for stellar and binary evolution, making possible the
creation of entirely new stellar species.  One obvious modification to
standard stellar evolutionary tracks arises from collisions and
mergers between stars, and we focus on that here.  From this
perspective, the clusters listed by Portegies Zwart et al.~(2004b;
Table 1) represent an ideal combination of properties, having ages of
less than a few million years and relaxation times of less than a few
tens of millions of years.  In these clusters, dynamical evolution,
traditionally regarded as a ``slow'' process, actually occurs much
more rapidly than the stellar evolution of even the most massive
stars.  In fact, cluster dynamics controls the early phases of these
stars' lives.

Portegies Zwart et al.~(2004b) are primarily concerned with the
lifetimes and global structural evolution of young dense clusters in
the vicinity of the Galactic center.  In this paper we consider mainly
the stellar evolutionary aspects of life in such an extreme
environment.  We start by investigating the circumstances under which
collisions are likely to occur, and how a cluster might find its way
into such a state.  We then present a scenario which may plausibly
lead to the formation of very massive stars and (perhaps)
intermediate-mass black holes (IMBHs) in sufficiently young, dense
systems.  Our results are based in large part on detailed $N$-body
simulations of model clusters.  Finally, we apply this scenario to
recent observations of the starburst galaxy M82.  In a companion
contribution, Baumgardt et al.~(2004) extend these ideas to the
subsequent evolution of a cluster containing a massive, compact
object.

%---------------------------------------------------------------------

\section{Stellar Collisions and Cluster Structure}

We are interested here in the possibility of runaway collisions
leading to ultramassive stars.  To appreciate the conditions under
which such runaways can occur, consider a massive object moving
through a field of background stars of total mass density $\rho$ and
velocity dispersion $v$.  We assume that the mass $M$ and radius $R$
of the object are large compared to the masses and radii of other
stars, and that all velocities are small enough that gravitational
focusing dominates the total cross section.  In that case, the
object's collision cross section is
\begin{equation}
 	\sigma \approx 2\pi G M R / v^2\,,
\end{equation}
nearly independent of the properties of the other stars.  The rate of
increase of the object's mass due to collisions is therefore
\begin{eqnarray}
	\frac{dM}{dt} &\approx& \rho\sigma v \nonumber\\
		      &\approx& 2\pi G M R \rho / v \nonumber\\
		      &=& 6\times10^{-11}
			\left(\frac{M}{M_\odot}\right)
			\left(\frac{R}{R_\odot}\right)\nonumber\\
		&~&~~~~~~~~~~~~~~~\times\,
			\left(\frac{\rho}{10^6\,M_\odot/{\rm pc}^3}\right)
			\left(\frac{v}{10\, {\rm km/s}}\right)^{-1}
			M_\odot/{\rm yr}\,.\ \ \ \ \ 
\end{eqnarray}
Thus, if the object initially has $M=100 M_\odot$ and $R=30 R_\odot$,
we fix $v$ at 10 km/s, and adopt a mass-radius relation $R\propto
M^{1/2}$, we find that in order for the object to accrete
$10^3M_\odot$ of material in 5 Myr (to form an IMBH within the typical
lifetime of a massive star), the local density must satisfy
\begin{equation}
 	\rho \ \ga\  5\times10^8\, M_\odot/{\rm pc}^3 \ \ =\ \
	 	\rho_{crit},\ \  {\rm say}\,.
\end{equation}
Such a density is much higher than the mean density of any known star
cluster, young or old.  For comparison, the average density of the
Arches cluster is $\sim$$6\times10^5\,M_\odot/{\rm pc}^3$, that of a
fairly compact globular cluster is $\sim$$10^4\,M_\odot/{\rm pc}^3$,
while even the most concentrated globular cluster cores have densities
$\la 10^{6-7}\, M_\odot/{\rm pc}^3$.

Might we be able to generate conditions more conducive to mergers by
assuming that a cluster is born very centrally concentrated
(e.g.~Portegies Zwart et al.~2004, Merritt et al.~2004)?  As a simple
limiting model of a very condensed cluster, consider the nearly
isothermal system of total mass $M_c$ and half-mass radius $r_h$,
described by the density profile
\begin{eqnarray}
	\rho(r) &=& \frac{M_c}{8\pi r_h r^2}\,,\\
	M(r)    &=& {\textstyle\frac12}M_c\left(\frac{r}{r_h}\right)\,,
\end{eqnarray}
for $0 \le r \le 2r_h$.  Densities exceeding $\rho_{crit}$ are found
for $r<r_{crit}$, where
\begin{eqnarray}
	r_{crit} &=& \sqrt{\frac{M_c}{8\pi r_h\rho_{crit}}} \nonumber\\
		 &=& 
			1.8\times10^{-3}
			\left(\frac{v}{10\,{\rm km/s}}\right)
%
%			2.5\times10^{-3}
%			\left(\frac{M_c}{10^5\,M_\odot}\right)^{1/2}
%			\left(\frac{r_h}{1 {\rm pc}}\right)^{-1/2} \ 
%
			{\rm pc}\,,
\end{eqnarray}
where $v = \sqrt{GM_c/2r_h} \sim 10$ km/s for all the clusters of
interest here.  However, the total mass contained within this radius
is just
\begin{equation}
	M_{crit} = 
			40 \left(\frac{v}{10\,{\rm km/s}}\right)^3
%
%			130 \left(\frac{M_c}{10^5\,M_\odot}\right)^{3/2}
%		       \left(\frac{r_h}{1 {\rm pc}}\right)^{-3/2}
%
			M_\odot\,.
\end{equation}
Given the highly optimistic assumptions needed to accrete even a small
fraction of this mass onto the original object, it is clear that, for
reasonable cluster parameters, there is far too little initial mass in
the high-density region to accomplish the task of forming a
$1000\,M_\odot$ object in the time available.

%---------------------------------------------------------------------

\section{Cluster Dynamics}

Thus it seems that collisions in a static cluster core cannot lead to
the formation of an ultramassive object.  However, it is well known
that cluster dynamical evolution can result in conditions much more
favorable for a runaway merger to occur.  Here we briefly describe the
relevant processes and their consequences.  (We note in passing that
essentially the same result could be achieved if there were
significant initial mass segregation in the cluster, but there is
currently no firm evidence to support this assumption.)

The evolution of a cluster is governed by its half-mass relaxation
time, the time scale on which two-body encounters transport energy
around the system:
\begin{equation}
	t_{rh}\ \approx\ \frac{0.14\, M_c^{1/2} r_h^{3/2}}
				{G^{1/2} m \ln\Lambda}
	      \ \approx\ 5\, \left(\frac{v}{10\,{\rm km/s}}\right)^3
			  \left(\frac{\bar{\rho}}
				{10^5\, M_\odot\,{\rm
					pc}^3}\right)^{-1}\ {\rm Myr}
%	      \ \approx\ \frac{v^3}{15\,G^2 m \bar{\rho} \ln\Lambda}
\label{trelax}
\end{equation}
(Heggie \& Hut 2003).  Here, $N$ is the number of stars in the system,
$m = M_c/N$ is the mean stellar mass, $\bar{\rho} = 3M_c/8\pi r_h^3$
is the mean cluster density, and $\ln\Lambda\sim \ln(0.1 N)\sim10$.
For an equal-mass system, the time scale for dynamical evolution---the
core collapse time---is about $15 t_{rh}$, too long to cause
significant structural change in a few million years as required here.
However, the presence of even a modest range in masses greatly
accelerates the process of core collapse (Spitzer 1987).  The time
scale for a star of mass $M$ to sink to the cluster center as
equipartition reduces its velocity is
\begin{equation}
      t_s(M) \ \sim\  \frac{m}{M}\, t_{rh}\,,\label{tseg}
\end{equation}
where we note that typical mass spectra (e.g.~Scalo 1986 or Kroupa
2001) have $m \sim 0.4$--0.6 $M_\odot$ ($0.5\,M_\odot$ is used in
Eq.~\ref{trelax}) and maximum mass $\sim$50--100 $M_\odot$.

From $N$-body simulations, Portegies Zwart \& McMillan (2002) find
that the most massive ($\ga 20 M_\odot$) stars segregate rapidly to
the cluster center, forming a dense stellar subcore on a time scale
$t_{cc}\sim0.2 t_{rh}$.  A central density increase of 2--3 orders of
magnitude is typical, boosting even a relatively low-density core into
the range where collisions become common, and greatly increasing the
reserve of raw material available to form a collision runaway.  The
collisions naturally involve the most massive stars in the cluster,
and the low relative velocities typical of these systems ensures that
the colliding stars merge with minimal mass loss (J.~Lombardi 2004,
private communication).

In systems having $t_{cc} \la 5$ Myr, corresponding to $t_{rh}\la 25$
Myr, essentially all the massive stars in the cluster can reach the
center before exploding as supernovae and hence participate in the
runaway process.  Using the parameters presented by Portegies Zwart et
al.~(2004b; Table 1), we find that the Arches and Westerlund I fall
into this category, while the Quintuplet, NGC 3603, and R\,136 all
come close.  In these cases, the maximum mass of the runaway is
limited primarily by the total number of massive stars in the
system---a few tenths of 1 percent of the total cluster mass.  In less
dense or more massive clusters, such as MGG-11, the relaxation time is
longer and only a fraction of the massive stars initially present in
the system can reach the center in the time available, but, as we will
see, their greater number may still ensure that a runaway can occur.

In small systems (containing less than a few tens of thousands of
solar masses), collision rates are significantly enhanced by the fact
that the massive object tends to form binaries, which are then
perturbed into eccentric orbits by encounters with other stars
(Portegies Zwart \& McMillan 2002).  Binary-induced mergers increase
the collision cross section, but they still require high central
densities before the (three-body) binary formation rates become
significant.  In larger systems, unbound collisions appear to be the
norm, and these have been observed in both direct N-body
(NBODY4/Starlab/treecode) and Monte-Carlo (G\"urkan et al.~2003)
cluster simulations.

%---------------------------------------------------------------------

\section{X-ray Sources and Dense Clusters in M82}

Recently a bright X-ray point source (M82 X-1) has been observed some
200 pc from the center of the starburst galaxy M82 (Matsumoto et
al.~2001; Kaaret et al.~2001).  With a luminosity exceeding $10^{41}$
erg/s, it is too bright to be an ordinary X-ray binary, while its
off-center location in the galaxy argues against its being a
supermassive black hole.  The luminosity is consistent with an
accreting compact object of at least 350 solar masses, raising the
intriguing possibility that it might be an IMBH.  The discovery of
$54.4\pm0.9$\,mHz quasi-periodic oscillations (Strohmayer et al.~2003)
supports this assertion.  Follow-up observations indicate that M82 X-1
is apparently located in the star cluster MGG-11, one of several
massive young clusters in the central region of M82.  This prompts the
obvious question: Could the observed X-ray source be the result of the
runaway collision process just outlined?

\begin{figure}[!htp]
\psfig{figure=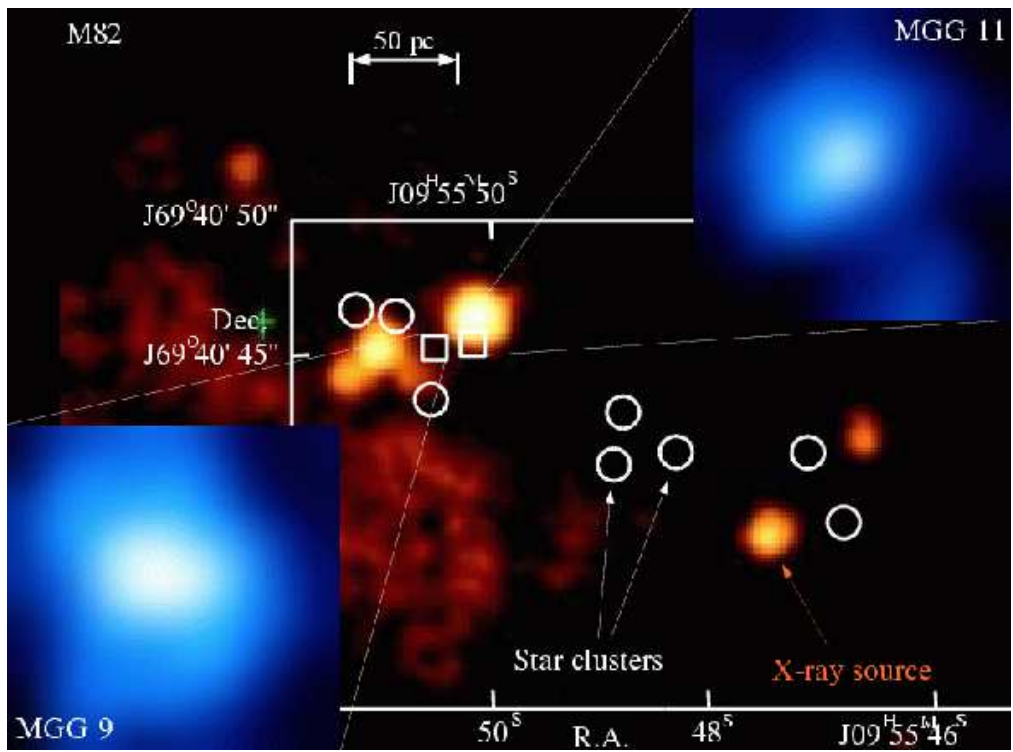,width=\columnwidth} 
\caption{Region of interest in M82.  The background X-ray image is
from Matsumoto et al.~(2001); M82 X-1 is near the center.  Star
clusters from Table 3 of McCrady et al.~(2003) are indicated by
circles and squares, the latter marking the positions of the two star
clusters MGG-9 and MGG-11.  Magnified near-infrared images of these
clusters from McCrady et al.~are presented at the upper right (MGG-11)
and lower left (MGG-9).}
\end{figure}

McCrady et al.~(2003) have made accurate measurements of the bulk
parameters of several clusters in the central regions of M82.  Figure
1 shows a composite X-ray (Matsumoto et al.) and near-infrared
(McCrady et al.) map of the few hundred parsecs around MGG-11.  The
relative positional accuracies of both the X-ray and the infrared
observations are better than 1 arcsecond.  However, the absolute
pointing accuracy is much poorer, for both telescopes.  Although
apparently off-center, the positions of the bright X-ray source and
the star cluster MGG-11 are in fact consistent with one another
(D.~Pooley 2004, private communication).  Curiously, the brighter,
more massive, and apparently coeval cluster MGG-9, lying just a few
arc seconds from MGG-11 in projection, shows no X-ray emission.

The clusters MGG-9 and MGG-11 have quite similar ages, in the range
7--12\,Myr (McCrady et al.~2003).  The line-of-sight velocity
dispersion of MGG-11 ($\sigma_r = 11.4\pm0.8$ km/s) is somewhat
smaller than that of MGG-9 ($\sigma_r = 15.9\pm0.8$ km/s).  Combining
these numbers with the projected half-light radii, 1.2 pc for MGG-11
and 2.6 pc for MGG-9, McCrady et al.~estimate total cluster masses of
$\sim (3.5\pm 0.7) \times 10^5 M_\odot$ for MGG-11, and about four
times higher for MGG-9.  The mean density of MGG-9 is just under half
that of MGG-11, raising a second question: Are such seemingly small
differences in cluster parameters sufficient to explain the presence
of a $>350 M_\odot$ IMBH in MGG-11 and the absence of a similar object
in MGG-9?

Portegies Zwart et al.~(2004a; PZBHMM) have addressed this issue using
detailed $N$-body simulations.  Starting with MGG-11, they first
demonstrate that IMBH formation is a natural outcome of that cluster's
dynamical evolution, and then go on to show that the same processes
would have failed to create a runaway in MGG-9.  Their calculations
were carried out using two independently developed $N$-body codes,
{\tt Starlab} (see Portegies Zwart et al.~2001) and {\tt NBODY4}
(Aarseth 1999, Baumgardt 2003).  Initial conditions for the model
clusters were chosen so that at the present time they have mass
functions, luminosities, half-mass radii and velocity dispersions in
agreement with the McCrady et al.~observations.

Since the initial and the current central densities of both clusters
are unknown, the concentration parameter $c$ (the logarithm of the
ratio of the tidal radius to the core radius) is treated as a free
parameter controlling the initial central density of the models.
PZBHMM find that, for $c > 2$ (which for ``King'' 1966 models is
equivalent to a dimensionless central potential $W_0 \ga 9$) the
MGG-11 models show runaway growth via repeated collisions.  The
mass-segregation time scale of a $50 M_\odot$ star in MGG-11 is
$t_s\sim 4$\,Myr (Eq.~\ref{tseg}).  Thus, massive stars in MGG-11 can
easily reach the center of the cluster before leaving the main
sequence.  Given the high central density of MGG-11, once those stars
have accumulated in the center, a runaway collision is inevitable,
leading to IMBHs with masses in the range 800--3000 $M_\odot$.  No
episode of runaway growth occurs in the MGG-11 models with $c < 2$,
nor in any of the MGG-9 simulations, regardless of initial
concentration.  In MGG-9, $t_s\ga 15 $\,Myr even for $100 M_\odot$
stars, so mass segregation cannot occur in the time available and no
runaway is seen.

\begin{figure}[!ht]
\psfig{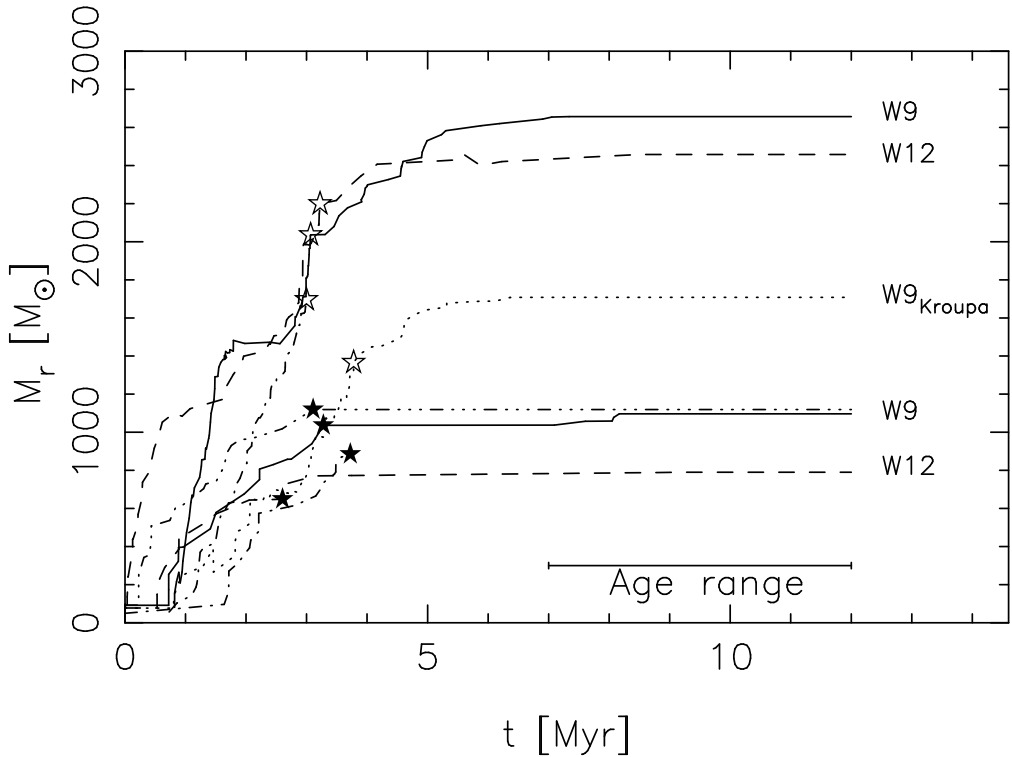} 
\caption[]{Growth in mass $M_r(t)$ of the collision runaway for some
of the simulations of PZBHMM, performed using {\tt Starlab} and {\tt
NBODY4}.  The choice of initial concentration parameter $W_0$ is
indicated.  The star symbols indicate the moment when the runaway
experiences a supernova, typically around 3\,Myr. Open and filled
stars indicate simulations performed with {\tt NBODY4} and {\tt
Starlab}, respectively.  The observed age range of MGG-11 and MGG-9 is
indicated by the horizontal bar near the bottom of the figure.  }
\label{fig:Mbh}
\end{figure}

Figure 2 presents a representative sample of results from a number of
simulations performed for a broad range of cluster parameters.  It
shows the growth in mass of the star that will ultimately become the
most massive object in the cluster.  Following detailed supernova
calculations by Heger et al.~(2003), stars having masses greater than
$260\,M_\odot$ are assumed to collapse to black holes without
significant mass loss.  The stellar evolution models for stars with
masses between 50 and $1000\,M_\odot$ are based on work by Stothers \&
Chin (1997) and Ishii et al.~(1999). The quantitative differences
between the simulations performed with {\tt Starlab} and those using
{\tt NBODY4} are due mainly to the different radii (and hence cross
sections) assumed for very massive stars in those two packages.

The solid and dashed curves in the figure show the runaway mass as a
function of time for a Salpeter (1955) IMF with a lower limit of $1
M_\odot$ and $c \approx 2.1$ ($W_0 = 9$) and $c \approx 2.7$ ($W_0=12$).
The dash-dotted curves are for two models with $W_0=9$ with an upper
limit to the IMF of $50 M_\odot$, instead of the standard $100
M_\odot$ used in the other calculations; these runs were terminated at
the moment the runaway star exploded as a supernova.  The
dash-3-dotted curve shows the result for $W_0=12$ with a Salpeter IMF,
with 10\% of the stars in primordial binaries---any tendency of these
systems to arrest core collapse is effectively offset by the larger
collision cross sections of the binary components.  Finally, the
dotted curve shows results for $W_0=9$ and a Kroupa (2001) IMF with a
minimum mass of $0.1 M_\odot$, in a simulation of 585,000 stars.

%---------------------------------------------------------------------

\section{Summary and Discussion}

Rapid mass segregation in a dense star cluster leads to an effective
core collapse on a time scale $\sim$$0.2 t_{rh}$ for typical initial
mass functions.  This in turn can lead to a runaway series of
collisions in the cluster core and the possible formation of a
$\sim$1000 $M_\odot$ IMBH there.  We therefore expect an association
between ultraluminous X-ray sources and the cores of dense young star
clusters.  A leading candidate for such an association is M82 X-1,
which appears to lie in the massive young cluster MGG-11.  On the
basis of $N$-body simulations and elementary considerations of the
time scale on which massive stars sink to the cluster center, we can
readily explain why MGG-11 might host an IMBH while its more luminous
neighbor MGG-9 does not.  High initial central concentrations are
required in order for this process to operate even in MGG-11, but we
note that all of the ``local'' clusters listed in Table 1 of Portegies
Zwart et al.~(2004b) are in fact very centrally condensed.

Of course, it must be conceded that next to nothing is known about the
detailed evolution and ultimate fate of stars hundreds or thousands of
times more massive than the Sun, so we should perhaps not take too
seriously the predictions of a $2000 M_\odot$ ``star'' in some of our
simulations.  Nevertheless, the simulations described here do make it
clear that the hearts of these dense stellar systems can easily
produce conditions suitable for repeated stellar collisions.  The
collision runaway at the center of such a system should be extremely
luminous and eminently observable during its short lifetime.
Observations of the cores of dense young star clusters in our Galaxy
and beyond may thus shed light on the structure and lifetimes of such
ultramassive stellar objects.

%---------------------------------------------------------------------

\end{document}